\pdfoutput=1 
\documentclass{appolb}

\usepackage{amsmath,amssymb,bbm,color}
\usepackage{epsfig}
\usepackage{euscript}
\usepackage{pslatex}
\usepackage{hyperref}
\usepackage{xcolor}
\usepackage{fancybox}
\usepackage{enumerate}
\usepackage{appendix}
\usepackage{comment} 

\numberwithin{equation}{section}
\numberwithin{figure}{section}
\numberwithin{footnote}{section}
\begin{document}

\title{
\vspace{-4.0cm}
\begin{flushright}
        {\small {\bf IFJPAN-IV-2014-1}}
\end{flushright}
\vspace{0.5cm}
{\bf\Large 
Monte Carlo study of NLO correction to QCD evolution kernel 
induced by the change of the factorization scale}
\thanks{
This work is partly supported by 
 the Polish National Science Centre grant DEC-2011/03/B/ST2/02632,
 the Polish National Science Centre grant UMO-2012/04/M/ST2/00240,
  the Research Executive Agency (REA) of the European Union 
  Grant PITN-GA-2010-264564 (LHCPhenoNet),
}}

\author{ \underline{M. Fabia\'nska}$^a$, S.\ Jadach$^b$
\address{$^a$ Marian Smoluchowski Institute of Physics, Jagiellonian University,\\
              ul. Reymonta 4, 30-059 Krak\'ow, Poland.}
\address{$^b$ Institute of Nuclear Physics, Polish Academy of Sciences,\\
              ul.\ Radzikowskiego 152, 31-342 Krak\'ow, Poland}
}


\maketitle

\begin{abstract}
The aim of the present study is to show that:
the redefinition of the factorization scale
$Q_i\to z_i Q_i$ in the ladder
can be traded exactly for the NLO correction 
to the LO evolution kernel,
$P(z)\to P(z)+(2C_F \alpha_S/\pi)\Delta(z)$
The above observation was done/exploited in the literature,
but the present study demonstrates how this phenomenon
is realized within the Markovian Monte Carlo parton shower --
hence it might be relevant in MC practice.
\end{abstract}



%

\section{ Introduction }

In the collinear factorization {\em the factorization scale} $Q$ limits
transverse phase space of all emitted particles.
Typical practical choices of $Q$ are:
virtuality of the emiter parton at the end of the multiple emission process,
maximum transverse momentum or maximum rapidity
of all emitted partons,
$\mu_F$ of the dimensional regularization,
total energy in the hard process $\sqrt{\hat{s}}$, etc.
Redefinition of the factorization scale may involve factor
$z$ being the relative loss of the energy of the emitter:
$Q\to z^\sigma Q$, $z=x_n/x_{0}$%
\footnote{Variable $x_i$ is 
  the standard lightcone (Bjorken) variable of the emitter parton after
  $i$-th emission, $i=1,2,3,...n$.},
$\sigma=\pm 1, \pm 2$. 
Many examples can be found in the literature, for instance:
(i) change from $\mu_F$ to virtuality in the hard process coefficient function
    ~\cite{Altarelli:1979ub},
(ii) change from time-like to space-like ladder 
  in the Curci-Furmanski-Petronzio (CFP)
  calculation of NLO kernels~\cite{Curci:1980uw},
(iii) change from angular- to kT-ordering
 in the modelling of low $x$ structure function
 by Catani-Ciafaloni-Fiorani-Marchesini (CCFM)~\cite{CCFM}.

The aim of the present study is to show that:
the redefinition of the factorization scale
$Q\to z^\sigma Q$ in the ladder
can be traded exactly for the NLO correction 
to the LO evolution kernel,
$P(z)\to P(z)+\sigma(2C_F \alpha_S/\pi)\Delta(z)$.
Without loss of generality,
in the numerical exercise we shall opt for $\sigma=1$.
As already said,
the above observation was already done/exploited in the literature.
Here, the above mechanism will be demonstrated {\em numerically},
in a form which can be useful
in the construction of the Monte Carlo parton shower
with the built in NLO evolution of the showers~\cite{Jadach:2010aa}.

\section{ Simplified DGLAP evolution in the Markovian Monte Carlo form}

For our numerical exercise we shall use
simplified DGLAP evolution in the Markovian Monte Carlo form.
We consider an incoming quark which emits gluons, before it enter hard process.
Its energy distribution $D(T,x)$
is a function of the evolution time $T=\ln Q^2$.
The DGLAP evolution equation ~\cite{DGLAP} reads%
\footnote{
We are using the following shothand notation
$
\big (f(\cdot)\otimes g(\cdot)\big)(x) 
\equiv \int_0^1 dzdy\; f(z)g(y) \delta(x-yz).
$
}:
\begin{equation}
\label{eq:evolution1}
\frac{\partial}{\partial T}D(T,x)=
\frac{2\alpha_s(T)}{\pi}
\int_x^1 \frac{dz}{z}P(z) D\Big(T,\frac{x}{z}\Big)=
\frac{2\alpha_s(T)}{\pi}
\Big[
P(\cdot) \otimes D(T,\cdot)
\Big](x)
\end{equation}
where $x$ is a part of initial energy 
(more precisely lightcone variable)
left after the emissions of a gluon from a quark.
The running QCD coupling constant is
$\alpha_s(T)=4\pi/(2\beta_0(T-\ln\Lambda_0))$~\cite{stirling-book}
where $\beta_0$ is that of ref.~\cite{GWP} 
and $\Lambda_0$ is the QCD scale parameter.
However, for the sake of simplicity we shall adopt constant $\alpha_s$
in the following numerical exercises.
The evolution kernel $P(z)$ is given by:
\begin{equation}
\label{eq:kernel}
\begin{split}
&P(z)=C_F
\bigg\{
        \frac{1+z^2}{2(1-z)_+}
        +\frac{3}{4}\delta(1-z)
\bigg\}
= -P^{\delta}(\epsilon)\delta(1-z)
       +P^{\theta}(\epsilon,z),
\end{split}
\end{equation}
where:
\begin{equation}
\label{eq:PdelThe}
\begin{split}
&P^{\theta}(\epsilon,z)=
\frac{C_F}{2}  \frac{1+z^2}{1-z}  \theta(1-z-\epsilon),
\quad
\\&
P^{\delta}(\epsilon)=
\int_0^1dz \;P^{\theta}(\epsilon,z)=
C_F \left[
        \ln \left(\frac{1}{\epsilon}\right)
       -\frac{3}{4}
    \right],
\end{split}
\end{equation}
$\epsilon\rightarrow 0$ is an infrared regulator and $C_F=\frac{3}{4}$ is the
colour-group factor. $P^{\delta}(\epsilon)$ is deliberately chosen to be
positive -- it is uniquely determined from the baryon number
conservation condition,
$
\int_0^1dz\;P(z)=0.
$

\begin{figure}
\begin{center}
\includegraphics[width=90mm]{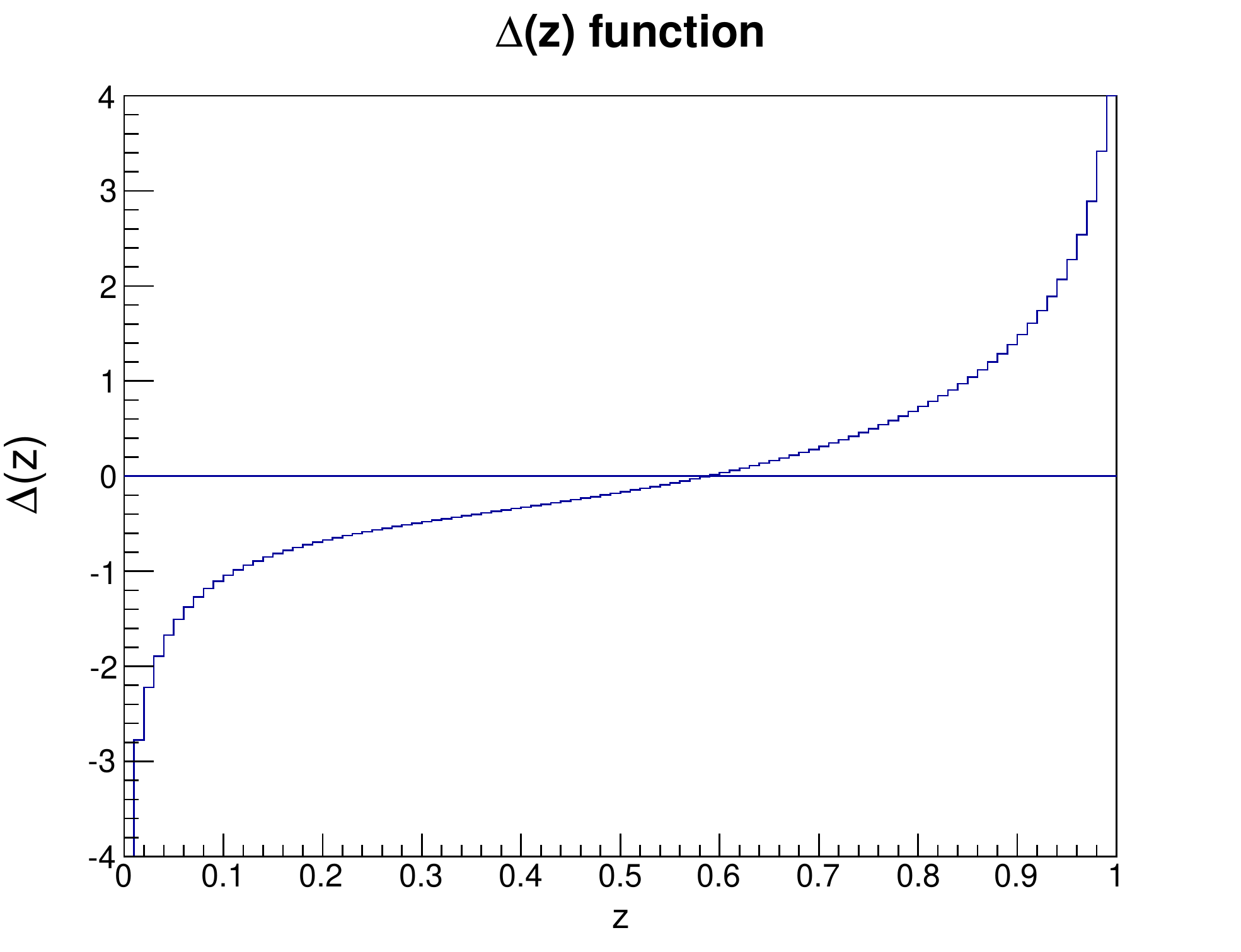}
\caption{The $\Delta$-function of Curci--Furmanski--Petronzio.}
\label{fig:deltaFig}
\end{center}
\end{figure}

The iteration of the above evolution equation leads to the following solution:
\begin{equation}
\begin{split}
\label{eq:iterative}
&D(T,x)=
e^{-\Phi(T,t_0)}D(t_0,x)+
\\&
+\sum_{n=1}^{\infty} \int_{t_0}^T \prod_{i=1}^n
    \bigg[
             dt_i \theta(t_i-t_{i-1})
    \bigg]
    e^{-\Phi(T,t_n)}\times
\\&
\times\prod_{j=1}^n
    \bigg[
       \frac{2\alpha_s(t_j)}{\pi}\; P^{\theta}(\epsilon,\cdot)
       e^{-\Phi(t_j,t_{j-1})}\otimes
    \bigg]
    D(t_0,\cdot)(x),
\end{split}
\end{equation}
where the Sudakov form-factor $\Phi(T,t_0)$ is given by
$
\Phi(T,t_0)=
\int_{t_0}^T dt' \;
\frac{2\alpha_s(t')}{\pi}\;
P^{\delta}(\epsilon).
$

On the other hand, the exact solution of the evolution equation
for $D(T,x)$ can be obtained
with high numerical precision from the Markovian Monte Carlo program.
The probability distribution
for generating single Markovian step forward,
that is generating the next $(t,x)$ starting 
from the previous $(t_0,x_0)$,
is given by:
\begin{equation}
\begin{split}
\label{eq:MarkovProb}
&p(t,x|t_0,x_0)=
\theta(t-t_0) 
\frac{2\alpha_s(t)}{\pi}
P^{\theta}
\left(
     \epsilon,\frac{x}{x_0}
\right)
e^{-\Phi(t,t_0)},\quad
\\&
\int_{t_0}^\infty dt 
\int_0^{x_0} dx\; 
p(t,x|t_0,x_0)=1.
\end{split}
\end{equation}
Our toy model Markovian Monte Carlo algorithm works as follows:
\begin{itemize}
\item
  $x_0$ is generated according to $D(x_0)=3(1-x_0)^2$,\;
  $\int_0^1 dx_0\; D(x_0) =1$.
\item
  $t_i=\ln(Q_i)$ and $z_i=\frac{x_i}{x_{i-1}}$ 
  are generated in a loop according to
  $p(t_i,x_i|t_{i-1},x_{i-1})$ for $i=1,2,3,...$
\item
  Markovian process (loop) is terminated at $i=N$,
  when $t_{N+1}>T$ for the first time.
\item
  The above procedure is repeated many times
  and the resulting distribution of the final $x=x_N$
  will be distributed according to $D(T,x)$
  being the solution of the evolution equation,
  see ref.~\cite{GolecBiernat:2006xw} for more details.
\end{itemize}

\subsection{ The $\Delta$-function of CFP }

In the perturbative QCD
the evolution kernel $P(z)$
is calculable order by order:
\begin{equation}
P(\alpha_s,z)=
P^{(0)}(z)
+\Big(\frac{\alpha_s}{2\pi}\Big)^1 P^{(1)}(z)
+\Big(\frac{\alpha_s}{2\pi}\Big)^2 P^{(2)}(z)+...,
\end{equation}
where $P^{(0)}(z)$, 
$P^{(1)}(z)$ and $P^{(2)}(z)$ are the leading (LO),
next-to-leading (NLO) and next-to-next-to-leading order (NNLO)
approximations respectively.
LO kernels are known since DGLAP works~\cite{DGLAP},
while NLO kernels were obtained directly from the Feynman diagrams 
in ref.~\cite{Curci:1980uw}.
In the same ref.~\cite{Curci:1980uw} it was noticed that
NLO corrections to the kernels for
the initial state ladder differ from the ones for final state by
$ \big(\frac{\alpha_s}{2\pi}\big) C_F^2 \Delta(z)$
where
\begin{equation}
\label{eq:CFP}
C_F^2 \Delta(z)=
\left[
  P^{(0)}(\cdot) \otimes \left(\ln(\cdot) \;P^{(0)}(\cdot)\right)
\right](z)
\end{equation}
and the LO kernel $P^{(0)}(z)=P(z)$ is that of eq.~(\ref{eq:kernel}).
The above $\Delta$-function is easily calculable:
\begin{equation}
\label{eq:nsimCFP}
\begin{split}
&\Delta(z)
= \int_0^1 dx\;
\bigg\{
    \frac{\theta(x>z)}{x}
    \frac{1+x^2}{2(1-x)}\;
    \ln(y) \frac{1+y^2}{2(1-y)}  \bigg|_{y=z/x}
\\&~~~~~~~~~~~~~~~~~~~~~~~~~   
-\frac{1+x^2}{2(1-x)}\;
    \ln(y) \frac{1+y^2}{2(1-y)}\bigg|_{y=z}
\bigg\}
\\&~~~~~~~
=\frac{1+z^2}{2(1-z)} \ln z 
\left[
    \ln\frac{(1-z)^2}{z}
   +\frac{3}{2} 
\right]
+\frac{1+z}{8} \ln^2 z
-\frac{1-z}{4} \ln z,
\end{split}
\end{equation}
This function is visualised in {\it Figure \ref{fig:deltaFig}}.
It obeys the sum rule $ \int_0^1 dz\;\Delta(z)=0$
due to $\int_0^1 P^{(0)}(z) dz=0$.

\section{ $\Delta$-function of CFP in the framework of Markovian MC}

In the following we are going to show with the help of the Markovian Monte Carlo
that the change of the time limit from $T$ to $(T+\ln x)$
induces a NLO correction to the
evolution kernel being $C_F^2 \frac{\alpha_s}{\pi}\Delta(z)$.

In the CFP work the $\Delta$-function is generated by the factor $x^{\epsilon}$,
see eq.~(2.61) in \cite{Curci:1980uw}.
Attributing the above factor to rescaling
of the factorization scale $\mu\to \mu/x$ and
defining $T=\ln\mu$, this results in the shift
$T\to T+\ln x$.

In our algorithm this change is realized in a slightly different way: 
by means of decreasing the value of the time limit $T$,
step by step, in every iteration of the loop:
after accepting a given step 
(by means of checking whether $t_{new}<T$  is satisfied)
we change the value of time limit $T$
at the $i$-the step in the following way:
\begin{equation}
T\rightarrow T+\ln(z_i).
\end{equation}

On the other hand, also within the Markovian MC,
instead of decreasing the time limit $T$, 
we add the NLO correction proportional to $\Delta$-function directly 
to the evolution kernel. 
More precisely it is done by means of correcting MC
events with the following MC weight:
\begin{equation}
\label{eq:weight}
w=\prod_i \frac {P^{(1)}(z_i)} {P^{(0)}(z_i)},
\end{equation} 
where $P^{(1)}(z_i)=P^{(0)}(z_i)+\lambda\Delta(z_i)$ and $\lambda = \frac{2C_F \alpha_s}{\pi}=0.100384$. Therefore the weight can be expressed as follows:
\begin{equation}
w=\prod_{i=1}^N 
 \bigg[
     1+ \lambda\;\Delta(z_i) \left( \frac{1+z_i^2}{2(1-z_i)} \right)^{-1}
 \bigg],\quad
\end{equation}
where $N$ is a number of emissions before the time limit $T$
is reached and $\Delta(z_i)$ is that of eq.~(\ref{eq:nsimCFP}).

\begin{figure}
\begin{center}
\includegraphics[width=90mm]{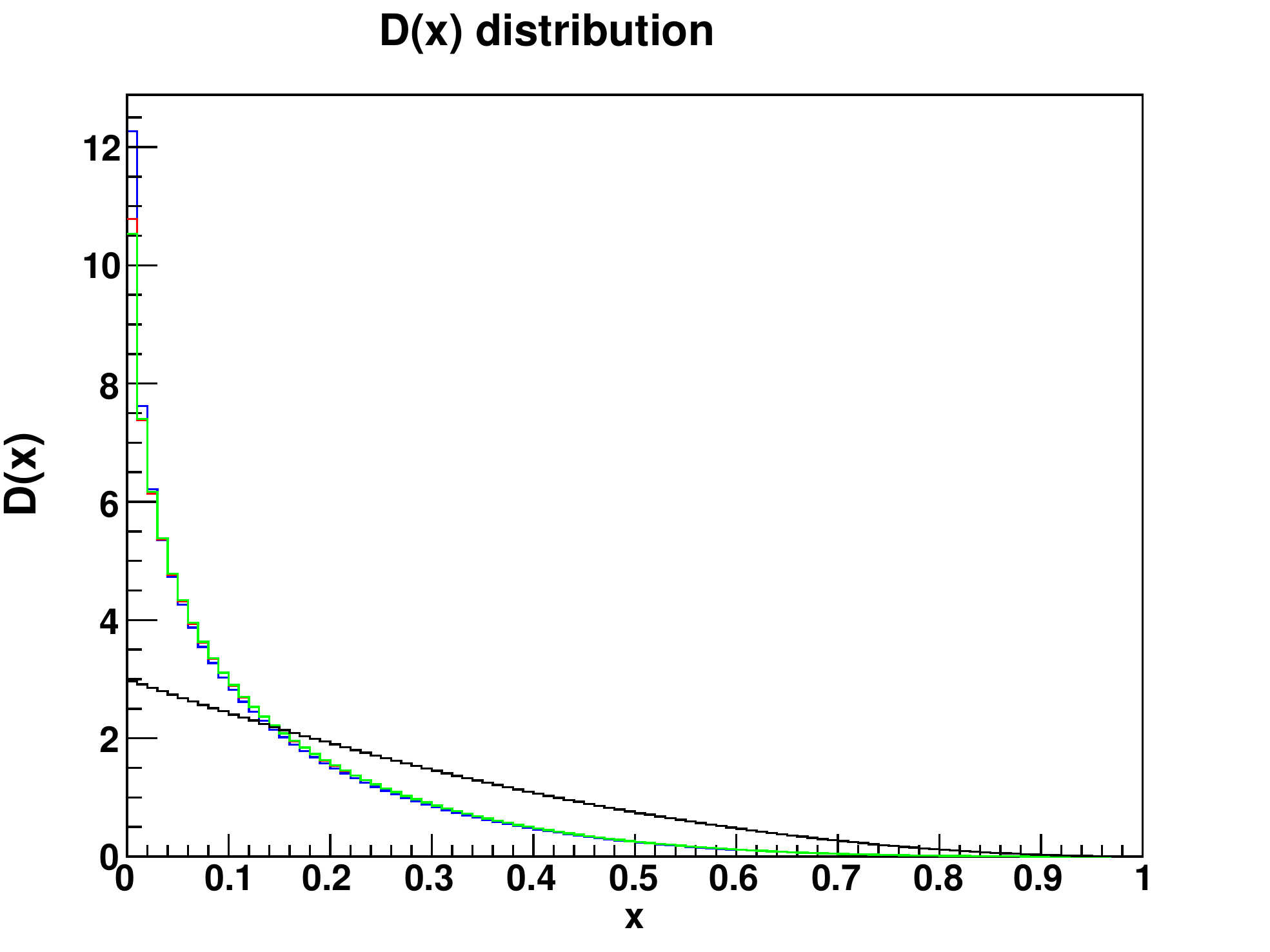}
\caption{Energy distributions $D(T,x)$:
(a) the one obtained by using the LO
approximation (blue), 
(b) the one obtained by decreasing the evolution time
limit (red), 
(c) the one obtained by correcting the LO kernel with the
$\Delta$-function (green) and 
(d) the initial energy distribution $D(x_0)$. 
They were generated using the following parameters:
$T=9.21034$ and $\epsilon=10^{-4}$. The distributions (b)-(d) coincide.}
\label{fig:DdistFig}
\end{center}
\end{figure}

\begin{figure}
\begin{center}
\includegraphics[width=90mm]{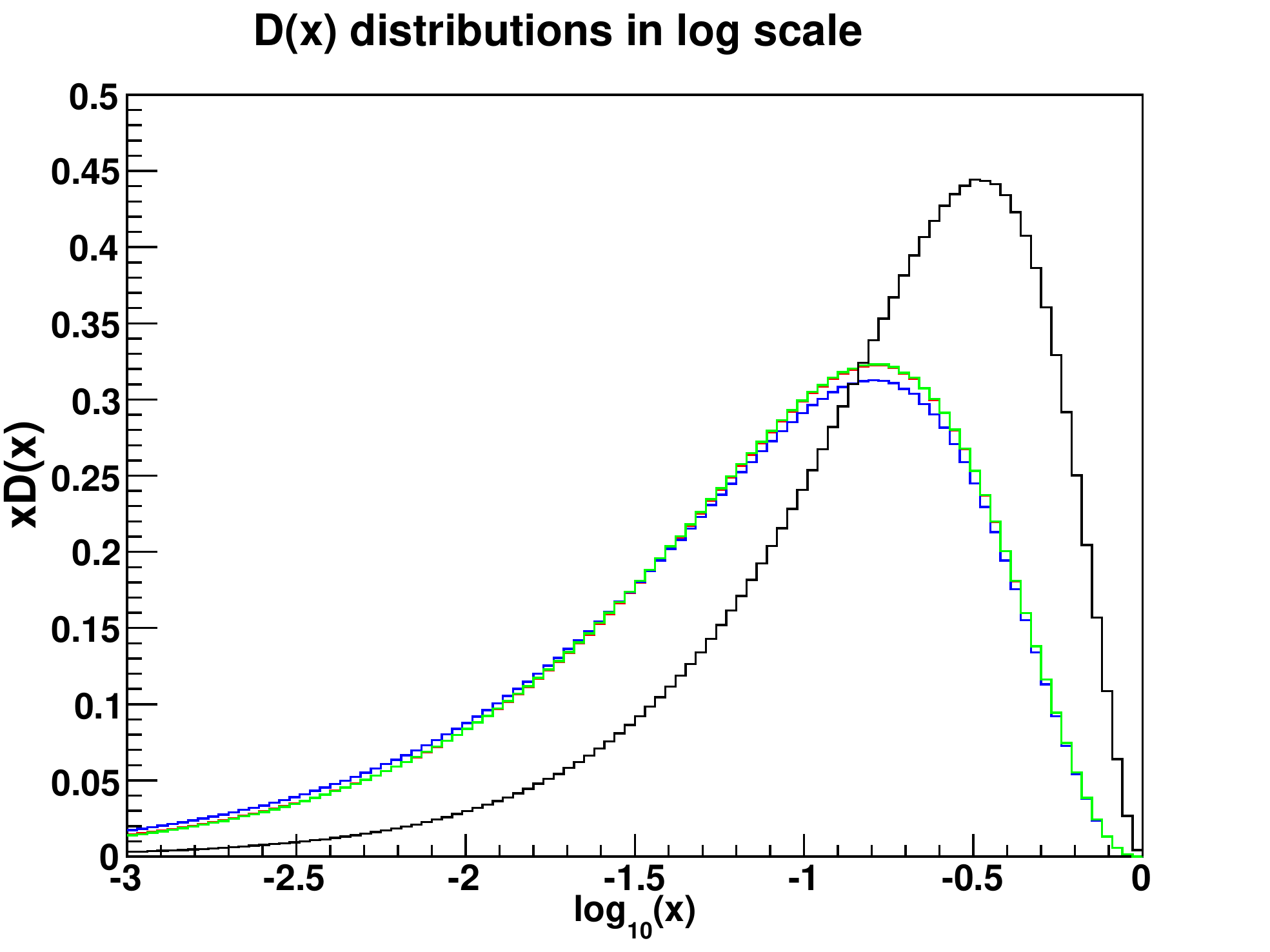}
\caption{ 
The same $xD(x)$ distributions as in Fig.~\protect\ref{fig:DdistFig}
plotted as functions of $log_{10}\;x$.
The parameters and the meaning of colours are also the same.
}
\label{fig:DdistLogFig}
\end{center}
\end{figure}

\subsection{ Numerical results }

Figure \ref{fig:DdistFig} shows various solutions $D(T,x)$ of the
evolution equation. 
The blue curve represents the solution
accurate up to LO. 
The red curve shows the distribution for the generation with
decreased time limit, 
while the green one shows the one obtained by adding the
$\Delta$-correction directly to the kernel
using eq.~(\ref{eq:weight}).
The black curve representing the 
initial distribution $D(t_0,x_0)$ is also shown.

It is clearly seen that red and green curves coincide, which confirms the statement
of Curci--Furmanski--Petronzio: decreasing the time limit has the same effect as
correcting the kernel with the $\Delta$-function. 

The differences between various curves
are better visible in {\it Figure \ref{fig:DdistLogFig}}, 
which shows the same
distributions multiplied by $x$ 
and plotted as a function of $log_{10}\;x$.

\begin{figure}
\begin{center}
{\includegraphics[width=90mm]{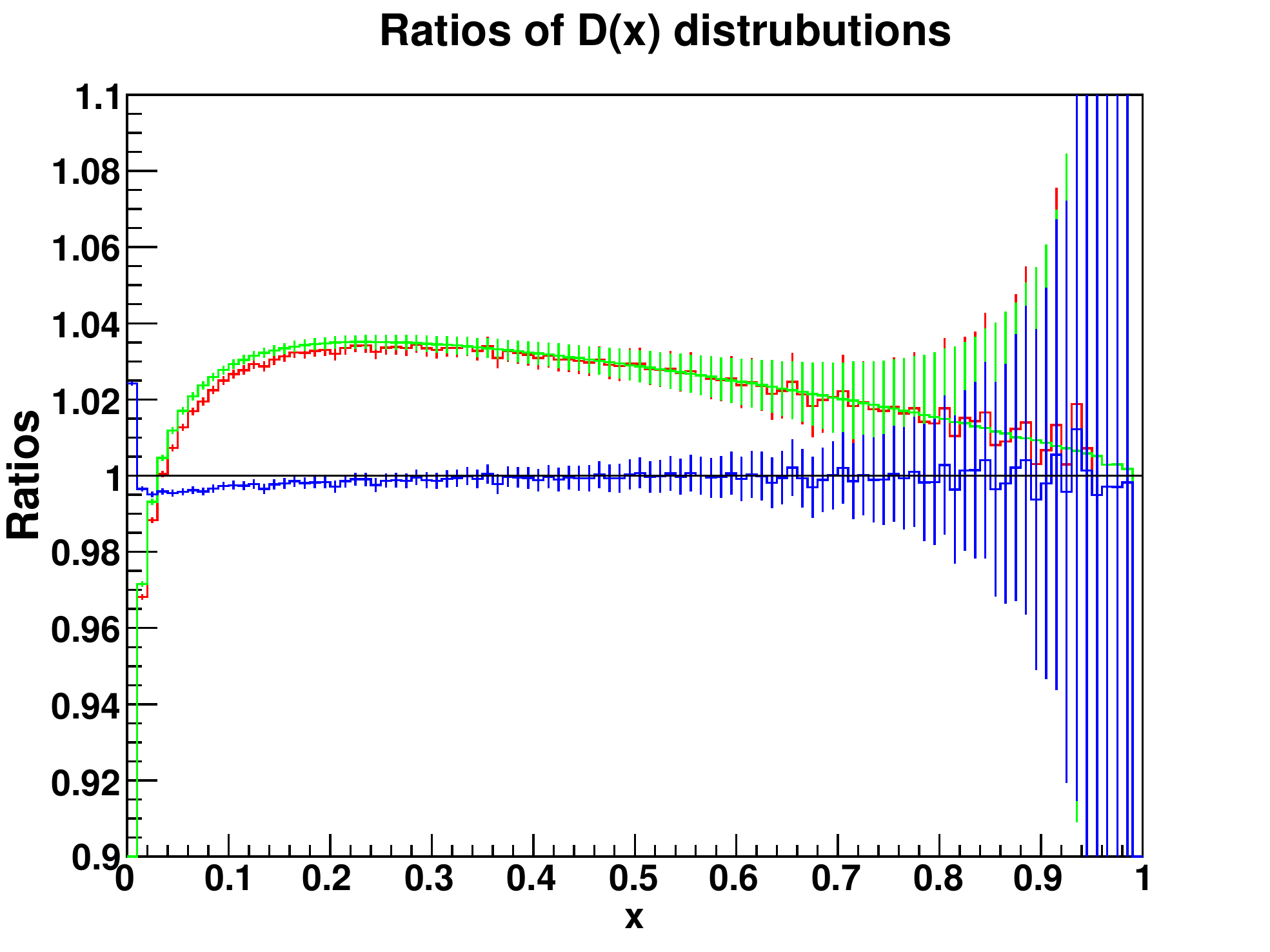}}
\caption{%
Ratios of $D(x)$ distributions: 
(b) divided by (a) (red), 
(c) divided (a) (green) and 
(b) divided by (c) (blue).
The notation (a), (b) and (c) and parameters
are the same as in Figure~\ref{fig:DdistFig}}.
\label{fig:ratials}
\end{center}
\end{figure}

\begin{figure}
\begin{center}
{\includegraphics[width=90mm]{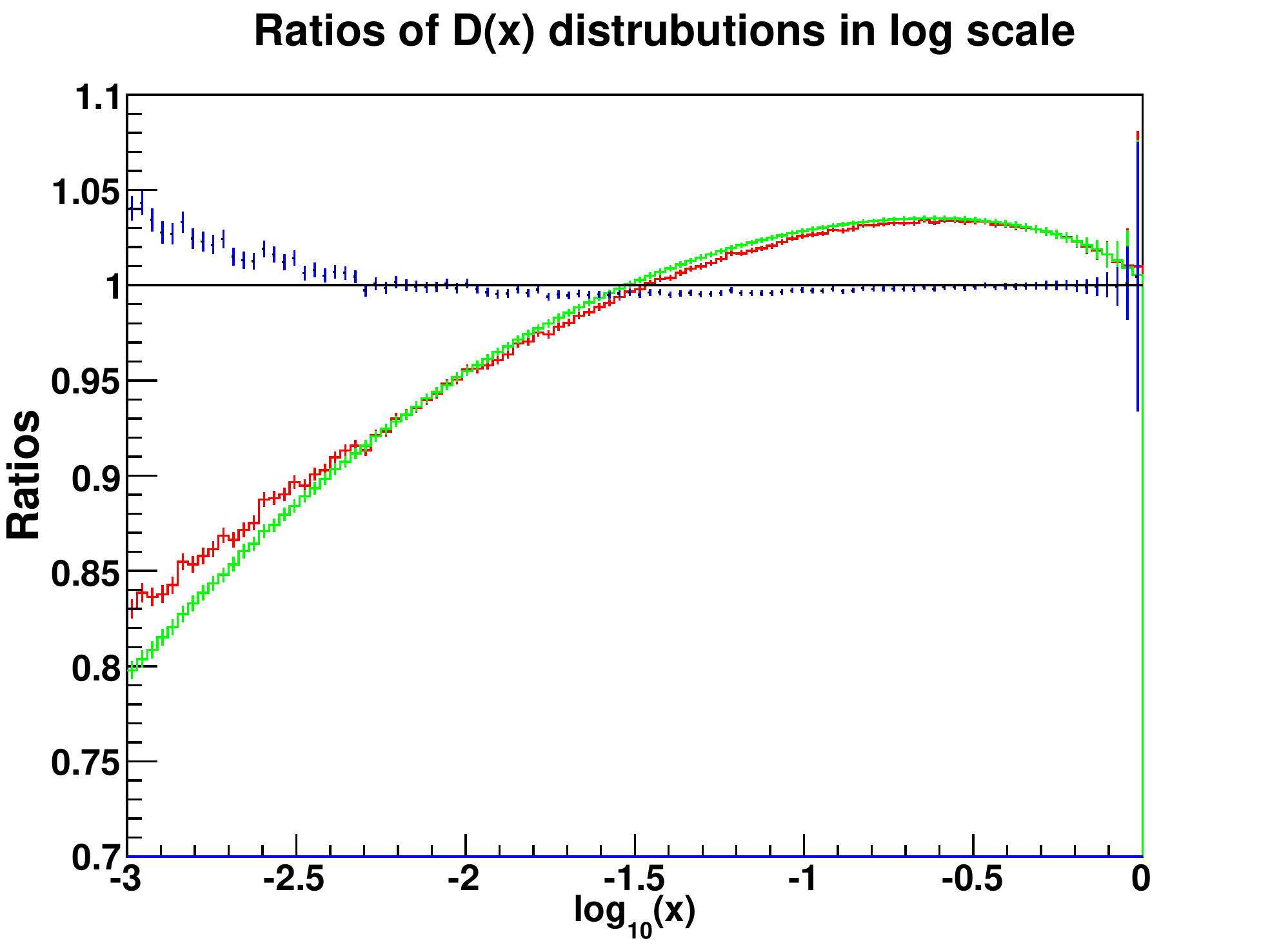}}
\caption{Ratios of $xD(x)$ distributions as functions of $log_{10}x$. 
The parameters and the meaning of colors are the same 
as in Figure \ref{fig:ratials}}.
\label{fig:ratialsLog}
\end{center}
\end{figure}

In order to see even better the differences between various resulting distribution
we plot in Figures \ref{fig:ratials} and \ref{fig:ratialsLog}
the ratios of the same distributions, 
once again as functions of $x$  and $log_{10}\;x$.
Now, the red curve represents the ratio of the solution obtained by
decreasing time limit and the one accurate up to the LO level. 
The green curve
shows the ratio of the solution obtained by using the direct $\Delta$-correction
to the kernel and the one obtained by using the LO approximation. 
Finally, the
blue curve represents the ratio of the distribution obtained by decreasing the 
time limit and the one with the direct $\Delta$-correction to the kernel. 
It is seen that the last ratio is close to one. 
Once more it indicates clearly our basic result that shifting
the evolution time limit by $\ln z$ (factorization scale by factor $z$) 
gives the same result as using the direct 
$\Delta$-correction to the kernel
in the way described by Curci--Furmanski--Petronzio~\cite{Curci:1980uw}.

We have checked that the slight
systematic difference between red and green curve 
in Figures \ref{fig:ratials} and \ref{fig:ratialsLog}
for small $x$ values results
from the fact that in the MC implementation shortening $T\to T +\ln(1/z_i)$
below the initial $t=0$ cannot be realized%
\footnote{ In the numerical exercise with $T\to T -ln(1/z_i)$
  and $\Delta \to -\Delta$ this discrepancy gets reduced.}.
Also, one has to keep in mind, that such a shortening evolution time limit
induces not only ${\cal O}(\alpha_s)$ contribution to the evolution kernel,
but also ${\cal O}(\alpha_s^2)$ term, which is not taken into account
in the present study.
Due to smallness of $\alpha_s$ the corresponding effect seems to be negligible.

Let us finally mention that all
plots and histograms presented in this sections have been obtained
using Monte Carlo software environment MCdevelop~\cite{Slawinska:2010jn}
and ROOT~\cite{Brun:1997pa} package.

\section{Summary}

The most important result presented here
is checking the equivalence of two methods of
implementing the $\Delta$-function of CFP in the Monte Carlo environment.
In the first method the evolution time range
was made shorter, step by step, after each iteration.
In the second method, the evolution time limit was kept fixed, but
the $\Delta$-function was added directly to the LO evolution kernel
as NLO correction,
by means of correcting generated events
with the help of a relevant MC weight.
Both methods have given the same results, within
the statistical error of the MC computations.
The small systematic difference between the results of both methods 
is the region of small $x$ values is well understood.


\providecommand{\href}[2]{#2}

\end{document}